\documentclass[a4paper,11pt]{article}
\usepackage{pos}
\usepackage{amsmath}
\usepackage[caption=false]{subfig}
\usepackage{bbm}
\usepackage{ulem}

\DeclareMathOperator{\Tr}{Tr}

\title{Infrared physics of the $\mathrm{SU}(2)$ Georgi-Glashow
  crossover transition }

\author{Lauri Niemi}
\author{Kari Rummukainen}
\author{Riikka Seppä}
\author*{David Weir}

\graphicspath{
  {./figures/}
}

\affiliation{Department of Physics and Helsinki Institute of Physics,\\
  P.O. Box 64, 00014 University of Helsinki, Finland}

\emailAdd{lauri.b.niemi@helsinki.fi}
\emailAdd{kari.rummukainen@helsinki.fi}
\emailAdd{riikka.seppa@helsinki.fi}
\emailAdd{david.weir@helsinki.fi}

\abstract{

We perform a lattice study of the phase transition in the
  $\mathrm{SU}(2)$ Georgi-Glashow model in three dimensions, where the
  gauge symmetry is broken to $\mathrm{U}(1)$ by the Higgs mechanism 
  and a photon-like state appears. Due to condensation of monopoles the photon 
  acquires a mass, which depends on the number density of the monopoles.
  We show that the monopole density can be renormalised on the
  lattice using gradient flow. Our preliminary results suggest that Polyakov's 
  semiclassical relation between the photon mass and the renormalised 
  monopole density is valid also at the nonperturbative level.
}

\FullConference{%
 The 38th International Symposium on Lattice Field Theory, LATTICE2021
  26th-30th July, 2021
  Zoom/Gather@Massachusetts Institute of Technology
}

\begin{document}
\maketitle

\section{Introduction}

Georgi-Glashow gauge theories, where scalar fields transform in the
adjoint representation of the gauge group, have a long history of
study in particle physics, notably as a grand unified theory when the
gauge group is, for example, $\mathrm{SU}(5)$~\cite{Georgi:1974sy}. It
also serves as a toy model of nonperturbative physics. In the simple
case of a single adjoint scalar and an $\mathrm{SU}(2)$ gauge group,
the system admits 't~Hooft-Polyakov monopole
solutions~\cite{tHooft:1974kcl,Polyakov:1974ek} and provides a
valuable testing ground for their behaviour. In this simple case, the
Higgs mechanism breaks the $\mathrm{SU}(2)$ gauge field to a
$\mathrm{U}(1)$ corresponding to a massless photon. However, monopole
condensation gives rise to a small nonzero photon mass
\cite{Polyakov:1976fu}.

Scalar fields transforming in the adjoint representation of
$\mathrm{SU}(2)$ are also of interest in theories of physics beyond
the Standard Model (BSM). In particular, they allow a more complicated
pattern of symmetry breaking phase transitions in the early universe
than the minimal Standard Model of the Higgs
doublet~\cite{Patel:2012pi}. Understanding how monopole-like
excitations may play a role in BSM phenomenology is crucial, as we go
beyond perturbation theory and take the first steps in probing these
models with lattice simulations~\cite{Niemi:2020hto}.

In this proceedings we explore the phase transition in the
$\mathrm{SU}(2)$ Georgi-Glashow model in three Euclidean dimensions,
which is the high temperature limit of the full $(3+1)$-dimensional
theory~\cite{Appelquist:1981vg}. This theory also serves as an
effective description of hot $\mathrm{SU}(2)$ gauge theory in the
deconfined phase~\cite{Kajantie:1997tt}.

Because of monopole condensation, the Higgs regime admits no massless
modes and is in fact analytically connected to the confining
regime~\cite{Kajantie:1997tt,Davis:2001mg}.  Semiclassically, the
photon mass squared $M_\gamma^2$ is expected to be proportional to the
monopole gas number density $n$~\cite{Polyakov:1976fu}. We explore
this relationship numerically at the nonperturbative level.

While measuring the nonzero photon mass $M_\gamma$ is relatively
straightforward, the monopole number density $n$ is a UV-divergent
quantity because of short-lived monopole-antimonopole pairs on the
lattice.  We show that the monopole number density can be renormalised
with the gradient flow~\cite{Luscher:2010iy}, so that it has a
well-behaved continuum limit. Our results support the hypothesis that
the semiclassical relation $M_\gamma^2 \propto n$ is valid for the
renormalised $n$.

\subsection{The model in continuum}

As outlined above, for our purposes the Georgi-Glashow model consists
of $\mathrm{SU}(2)$ gauge fields and a Higgs field $\phi$ in the
adjoint representation. By dimensional reduction, its infrared
behavior at high temperature is well described by a 3D theory with
temperature dependent parameters:
\begin{equation}
  S = \int \mathrm{d}^3x \; \left\{ \frac{1}{2} \mathrm{Tr}\, F_{ij} F_{ij} + \mathrm{Tr}\, [D_i,
    \phi]^2 + m^2_3(T) \mathrm{Tr}\, \phi^2 + \lambda_3(T) (\mathrm{Tr}\, \phi^2)^2
  \right\}.
\end{equation}
The same 3D action arises as the high-temperature limit of two-color
QCD, and of beyond the Standard Model theories involving electroweak
triplet scalars. The phase structure depends on two dimensionless ratios,
\begin{equation}
  x = \frac{\lambda_3}{g_3^2} \quad \text{and} \quad y =
  \frac{m^2_3}{g_3^4},
\end{equation}
where $g_3$ is the gauge coupling in 3D.

The monopoles give a mass to the photon-like
excitation. Semiclassically~\cite{Polyakov:1976fu}
\begin{equation}
  M_\gamma^2 \sim \frac{n}{\pi g_3^2}, \quad\quad n \sim
  \frac{m_W^{7/2}}{g_3} \exp\Big[-\frac{4\pi m_W}{g_3^2} f(\lambda_3 /
    g_3^2)\Big],
\label{eq:semiclassicalrelationship}
\end{equation}
where $n$ is a renormalised monopole number density that counts only widely
separated monopoles, and $f(z)$ an $\mathcal{O}(1)$ function.

\subsection{Lattice formulation}

Writing the $\mathrm{SU}(2)$ links as $U_{i}$, our lattice action reads
\begin{multline}
  S = \beta \sum_{x, i<j} \Big(1 - \frac{1}{2} \Tr U_{ij}(x) \Big) + 2 a
  \sum_{x,i} \Big( \mathrm{Tr}\, \phi(x)^2 - \mathrm{Tr}\, \phi(x)
  U_i(x)\phi(x+i) U^\dagger_i(x) \Big) \\
  + a^3 \sum_x \Big( m^2_L
  \mathrm{Tr}\, \phi^2 + \lambda_3 (\mathrm{Tr}\, \phi^2)^2 \Big) ,
  \label{eq:action}
\end{multline}
where $U_{ij}$ is the plaquette and $\beta = \frac{4}{a g_3^2}$. For
small $a$ the mass parameter is related to the continuum one by $m_L^2
= m_3^2 + \delta m^2$, where the counterterm is given in
\cite{Laine:1995np}. Letting $\Pi_+ = \frac{1}{2} (\mathbbm{1} + \phi
/ \sqrt{\phi^2})$, a $\mathrm{U}(1)$ link variable can be projected
out as
\begin{equation}
  u_i(x) = \Pi_+(x) U_i(x) \Pi_+(x+\hat\imath).
\end{equation}
The associated $\mathrm{U}(1)$ field-strength tensor is
\begin{equation}
  \alpha_{ij}(x) = \frac{2}{g_3 a^{1/2}} \arg \mathrm{Tr}\, u_i(x) u_j(x+\hat{\imath}) u^\dagger_i(x+\hat{\jmath}) u^\dagger_j(x).
\end{equation}
This in turn provides a meaningful definition
of the magnetic field $B_i$ on the lattice~\cite{Davis:2000kv},
\begin{equation}
  B_i(x) = \epsilon_{ijk} \alpha_{jk}(x).
\end{equation}
The magnetic monopole number density is
\begin{equation}
  n = \frac{1}{V} \frac{g_3 a^{1/2}}{4\pi} \sum_{x,i} \Big| B_i(x+\hat\imath)
  - B_i(x) \Big|.
\end{equation}
However, it is ultraviolet divergent due to short-lived
monopole-antimonopole pairs. We renormalise it with gradient flow of
the fields. If we parametrise the links as $U_i(x) = \exp \left[i
  \theta_i^c(x) \sigma^c / 2 \right]$, the gradient flow for this
model in terms of dimensionless flow time $\tau$ is
  \begin{align}
  \label{eq:grad-flow-1}
  \frac{\partial U_i(x)}{\partial\tau} &= -i g_3^2 a \frac{\sigma^c}{2} \frac{\partial S}{\partial \theta_i^c(x)} U_i(x) \\
  \label{eq:grad-flow-2}
  \frac{\partial \phi^c(x)}{\partial \tau} &= - a^{-1} \frac{\partial S}{\partial \phi^c(x)}.
  \end{align}
The gradient flow transforms fields towards saddle point
configurations of the action, removing ultraviolet fluctuations
through smoothing~\cite{Luscher:2010iy}. The smoothing radius $\xi$ in
3D is related to the flow time by $\xi = \sqrt{6\tau}a$.

At small $a$, the Higgs mass parameter $m_L^2$ requires a large
negative counterterm to preserve connection with continuum physics
\cite{Laine:1995np}. However, the gradient flow removes UV
fluctuations, and effectively reduces the need for the counterterm.
Applying the flow to the lattice action (\ref{eq:action}) with the
counterterm included in $m_L^2$ would drive the Higgs field to larger
values as the gradient flow proceeds. To avoid this, we choose
  the simplest option and explicitly remove the counterterm for the
gradient flow and use the continuum mass $m_3^2 = y g_3^4$
instead. This constitutes a choice of the renormalisation scheme,
giving us well-behaved results for the monopole number density.

\section{Results}

We focus on $x = 0.35$, for which the confinement-Higgs transition is
of the crossover type in this model~\cite{Kajantie:1997tt}. We use the
heatbath and over-relaxation algorithms to update the gauge and Higgs
fields respectively, and generate field configurations in the
canonical ensemble. An additional Metropolis step is added to the
Higgs updates for ergodicity.

\begin{figure}
  \centering
  \includegraphics[width=0.6\textwidth]{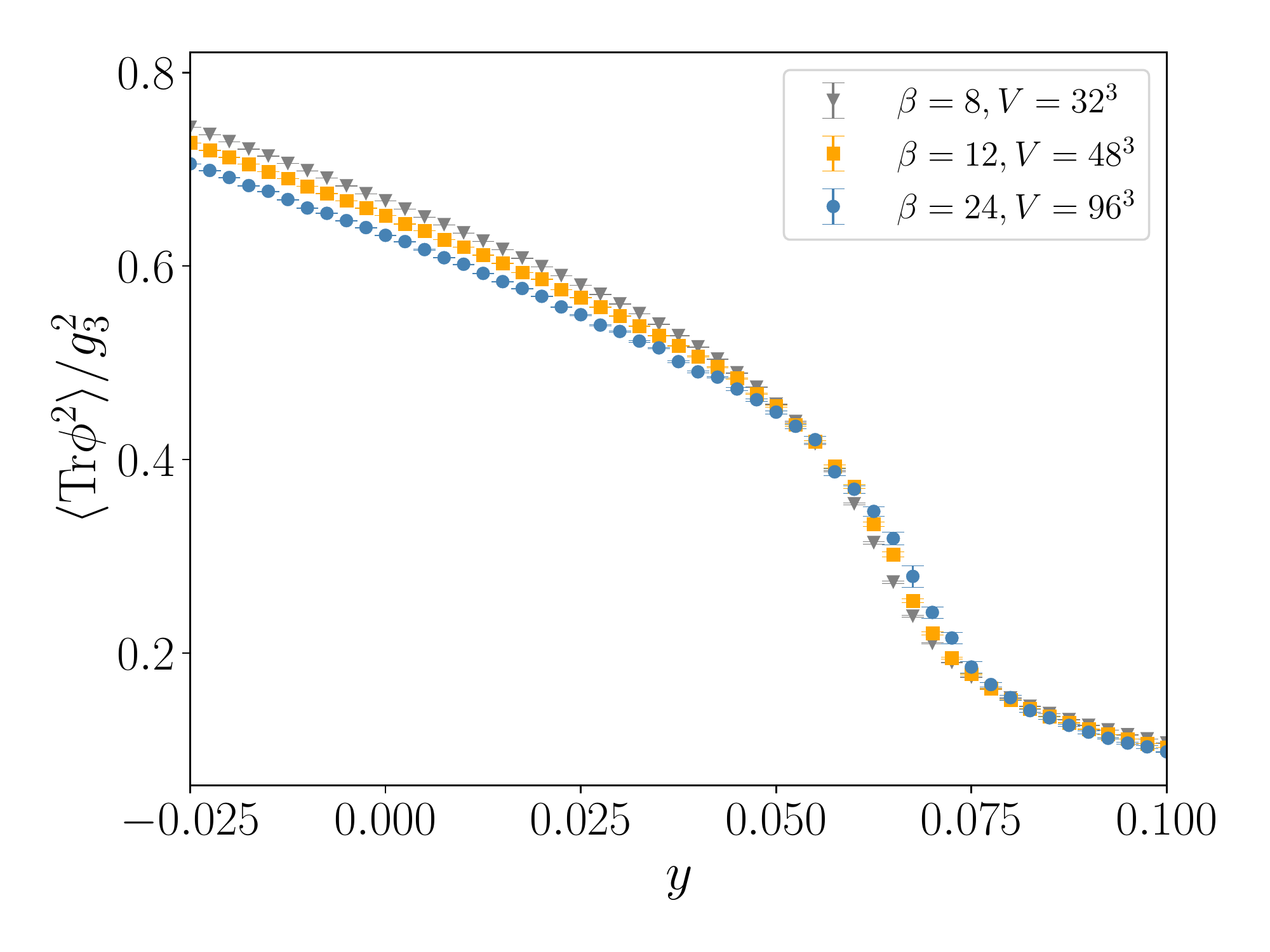}
  \caption{Higgs field expectation value for various lattice spacings,
    converted to continuum $\overline{\text{MS}}$. \label{fig:condensate}}
\end{figure}
  
In Fig.~\ref{fig:condensate}, we plot the Higgs field expectation
value through the crossover, noting that the renormalised expectation
value has a clear continuum limit. The lattice cutoff effects vanish
linearly in $1/\beta$~\cite{Laine:1995np}.  This allows us to identify
the approximate location of the pseudocritical point.

\subsection{Renormalised monopole density from gradient flow}

\begin{figure}
  \centering
  \subfloat[$\xi = 0.69a$]{
    \includegraphics[trim=257 96 154 346,
      clip,width=.32\textwidth]{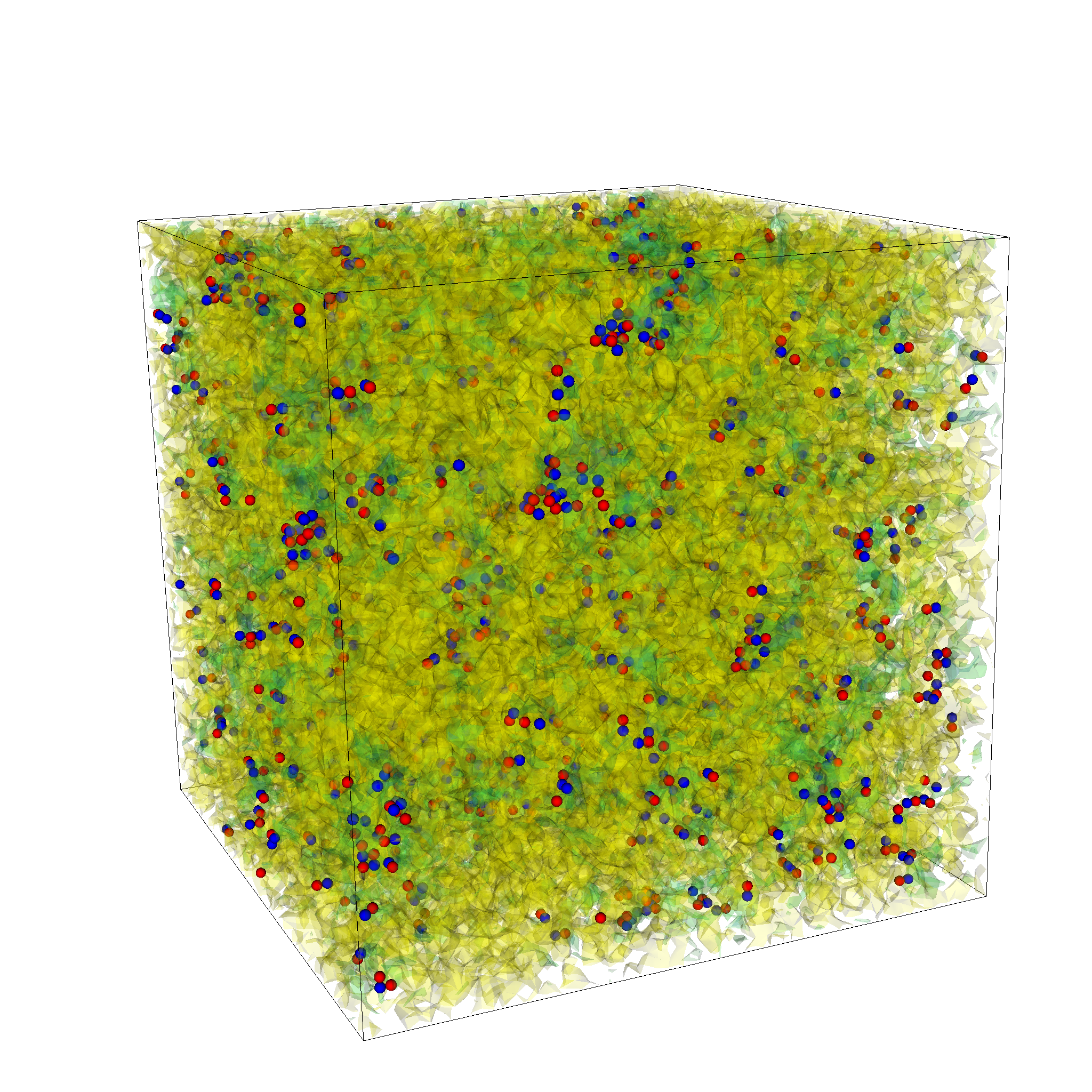}
  }\hfill
  \subfloat[$\xi = 1.55a$]{\includegraphics[trim=257 96 154 346,
      clip,width=.32\textwidth]{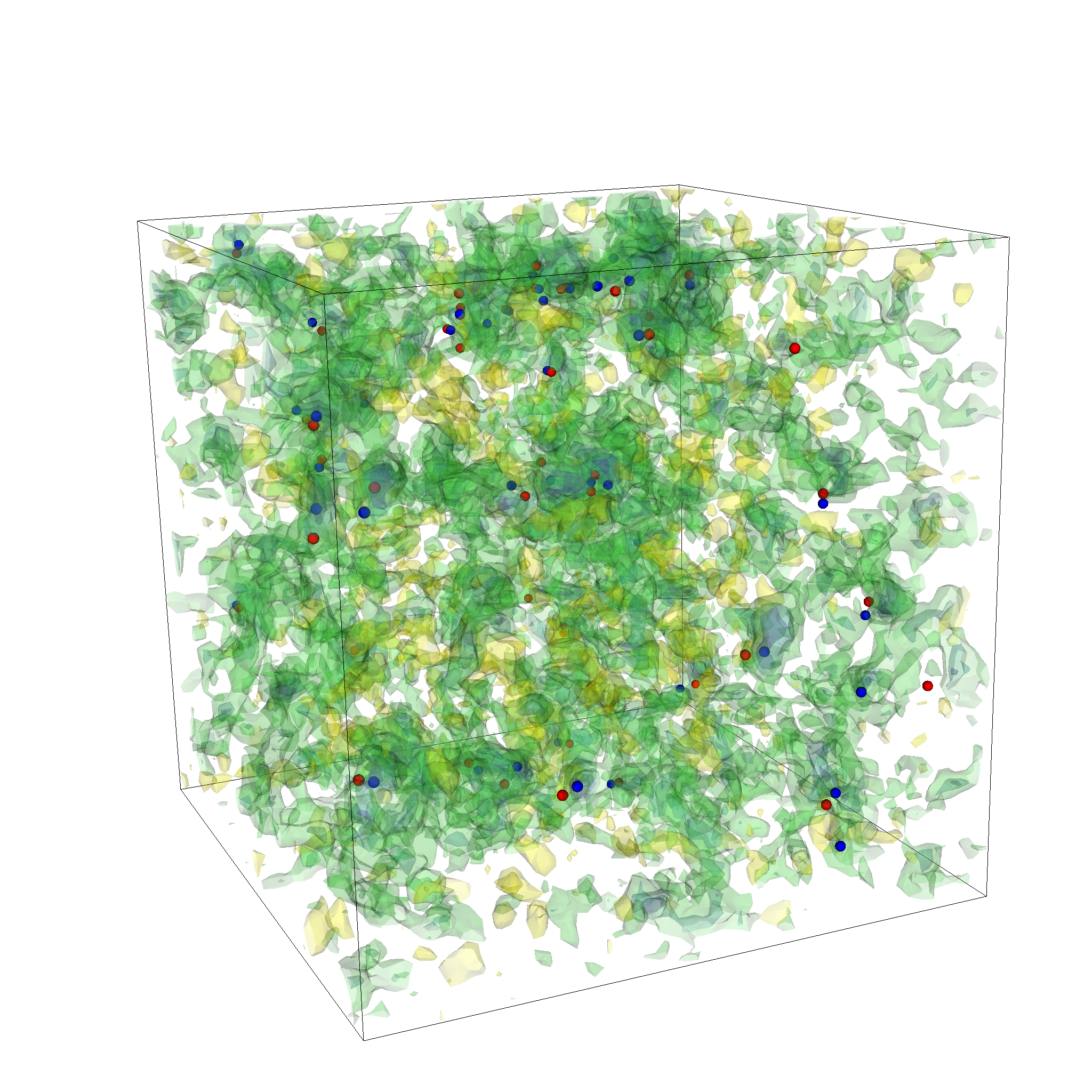}
  }\hfill
  \subfloat[$\xi = 3.46a$]{\includegraphics[trim=257 96 154 346,
      clip,width=.32\textwidth]{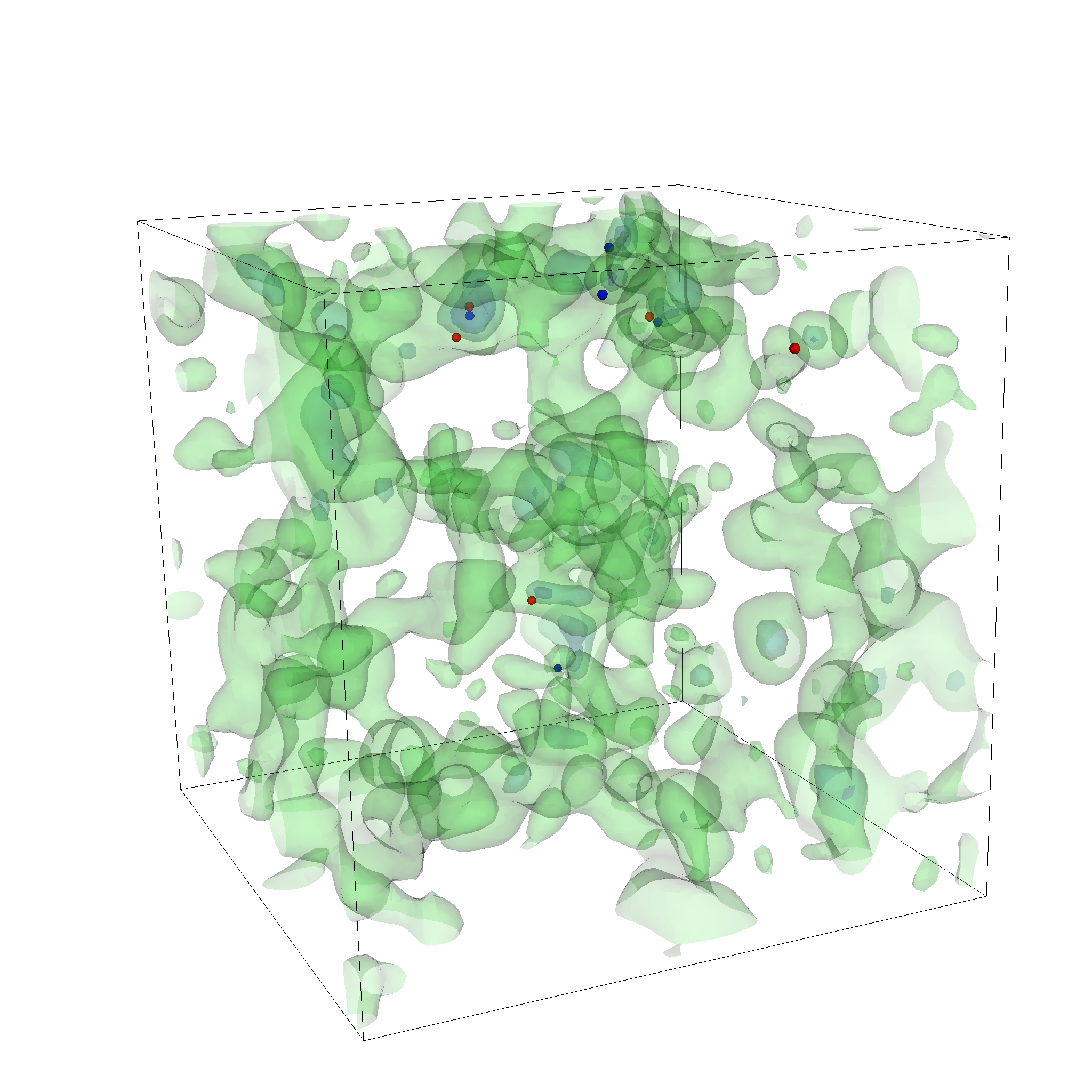}
  }
  \caption{Snapshots of the system at different stages of gradient
    flow, parameterised by the smoothing radius $\xi$ in terms of the
    lattice spacing $a$. Isosurfaces of the Higgs field are shown in
    shades of green, and (anti)monopoles are shown in red
    (blue).\label{fig:snapshots}}
\end{figure}

\begin{figure}
  \centering
  \includegraphics[width=0.6\textwidth]{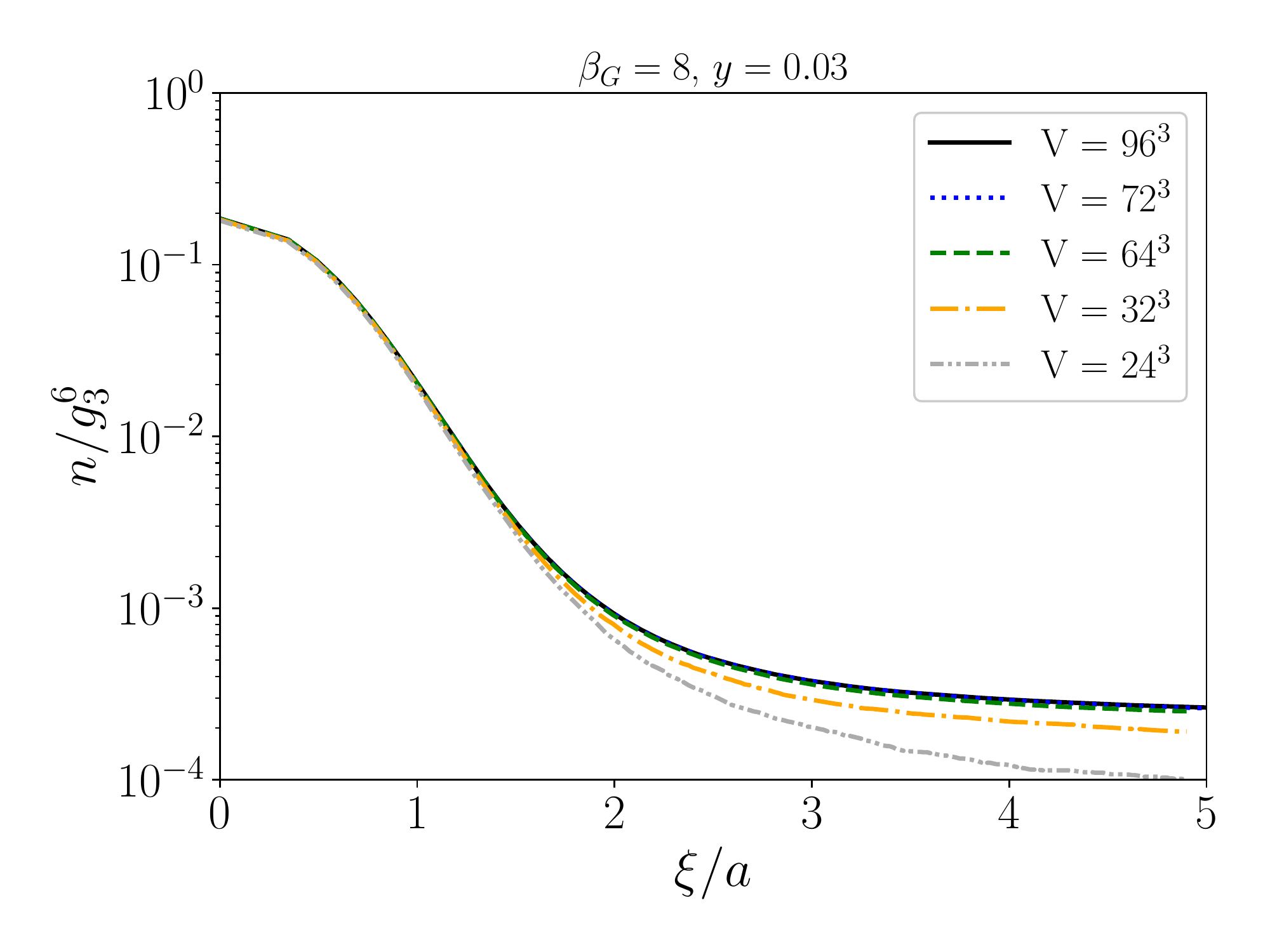}
  \caption{Number density of monopoles $n$ measured as a function of
    flow time $\xi$, at fixed $\beta$. At small volumes, it seems that
    all the monopoles annihilate efficiently. As the volume is
    increased, monopoles experience screening, and one can infer an
    infinite-volume limit. Note the semilogarithmic
    scale.\label{fig:flowtime}}
\end{figure}

As discussed above, we perform gradient flow (see
Fig.~\ref{fig:snapshots}) of the fields on the lattice in order to
measure a renormalised monopole number density and remove ultraviolet
monopole-antimonopole pairs. In Fig.~\ref{fig:flowtime}, we show the
effect of the gradient flow on the number density of monopoles. We
note that the monopoles are screened from each other at long
distances, but the lattice needs to be relatively large to capture
this effect~\cite{Davis:2001mg}.

\begin{figure}
  \centering
  \includegraphics[width=0.6\textwidth]{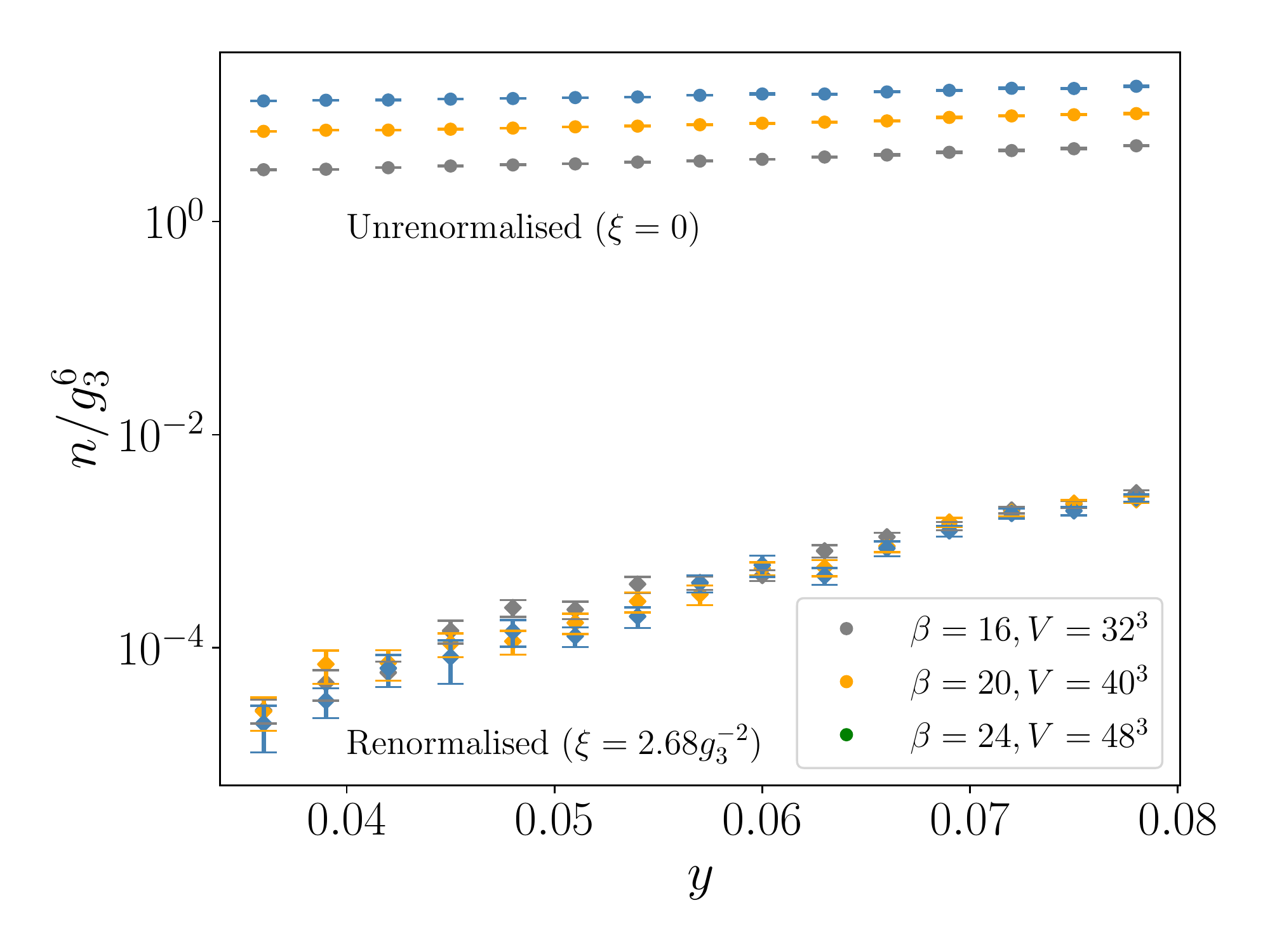}
  \caption{Unrenormalised and renormalised monopole densities at
    different spacings. The renormalised results, taken at a cooling
    time of $\xi=2.68 g_3^{-2}$, indicate a well-behaved $a\rightarrow
    0$ limit. \label{fig:renormalised_density}}
\end{figure}

The monopoles become heavier as the Higgs condensate grows, and the
monopole density drops rapidly. In
Fig.~\ref{fig:renormalised_density}, we plot both the unrenormalised
and renormalised number densities of monopoles for a number of lattice
spacings. These results strongly indicate that the renormalised
monopole density has a well-defined continuum limit. However, judging
from Fig.~\ref{fig:flowtime}, the lattices used here are likely too
small to fully see the screening of monopoles in the infrared. We do
not expect this limitation to qualitatively affect the $a\rightarrow
0$ properties of $n$. A detailed investigation of finite-volume
effects will be presented in an upcoming publication.

\subsection{Photon mass}

We measure the photon mass using blocked correlators at non-vanishing
momentum \cite{Kajantie:1997tt}. Within the accuracy of the
measurement the photon is massless deep in the Higgs regime, where the
monopole gas is dilute.  This agrees with the earlier lattice study
of~\cite{Hart:1996ac}.  The photon mass squared is approximately
proportional to the monopole density, in accordance with semiclassical
expectations. Furthermore, this relationship holds in the crossover
region where perturbation theory cannot be relied upon. The
proportionality constant depends on how much gradient flow cooling is
applied; in other words, it depends on the renormalisation scale.

\begin{figure}
  \centering
  \includegraphics[width=0.45\textwidth]{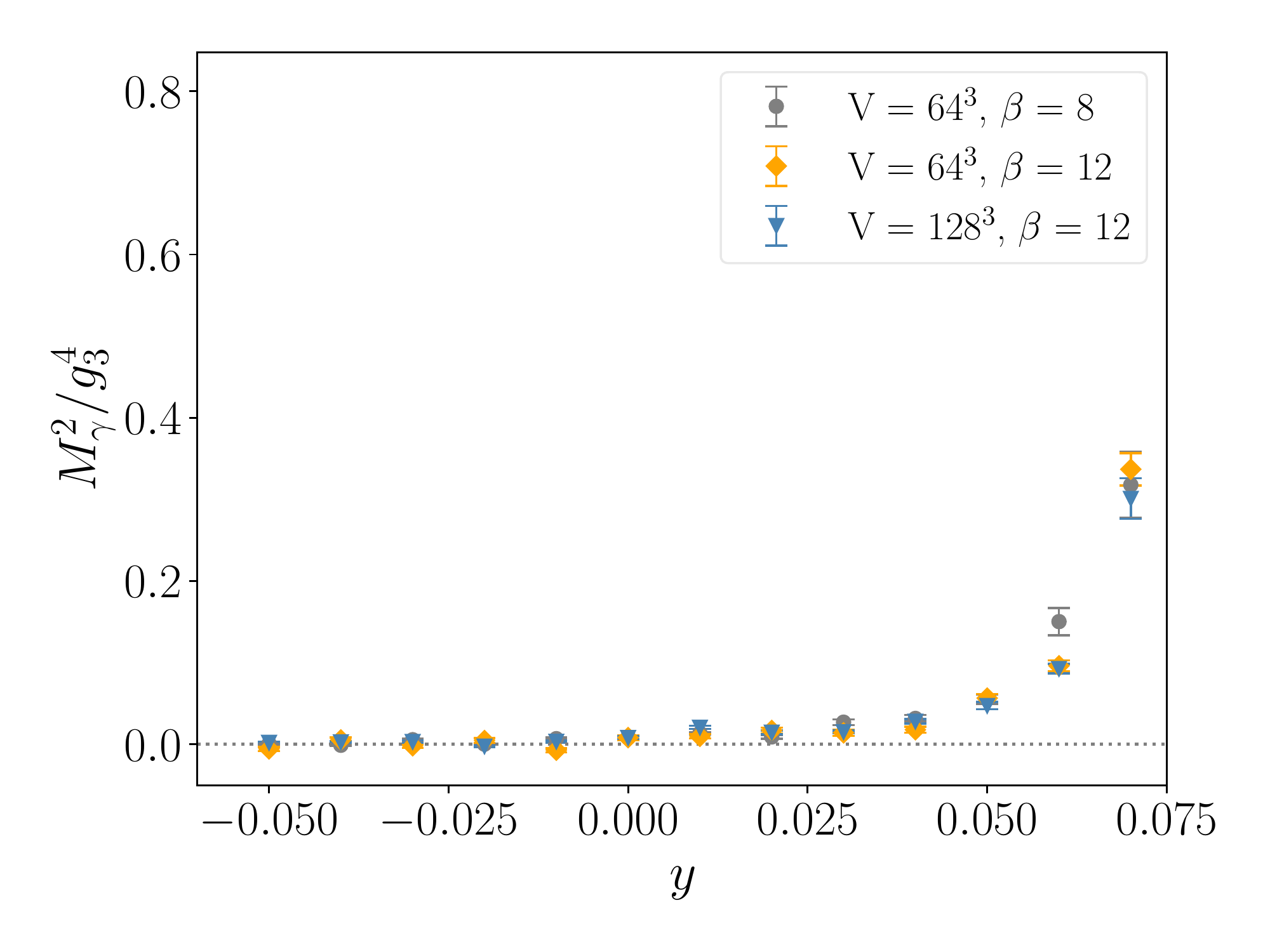}
  \includegraphics[width=0.45\textwidth]{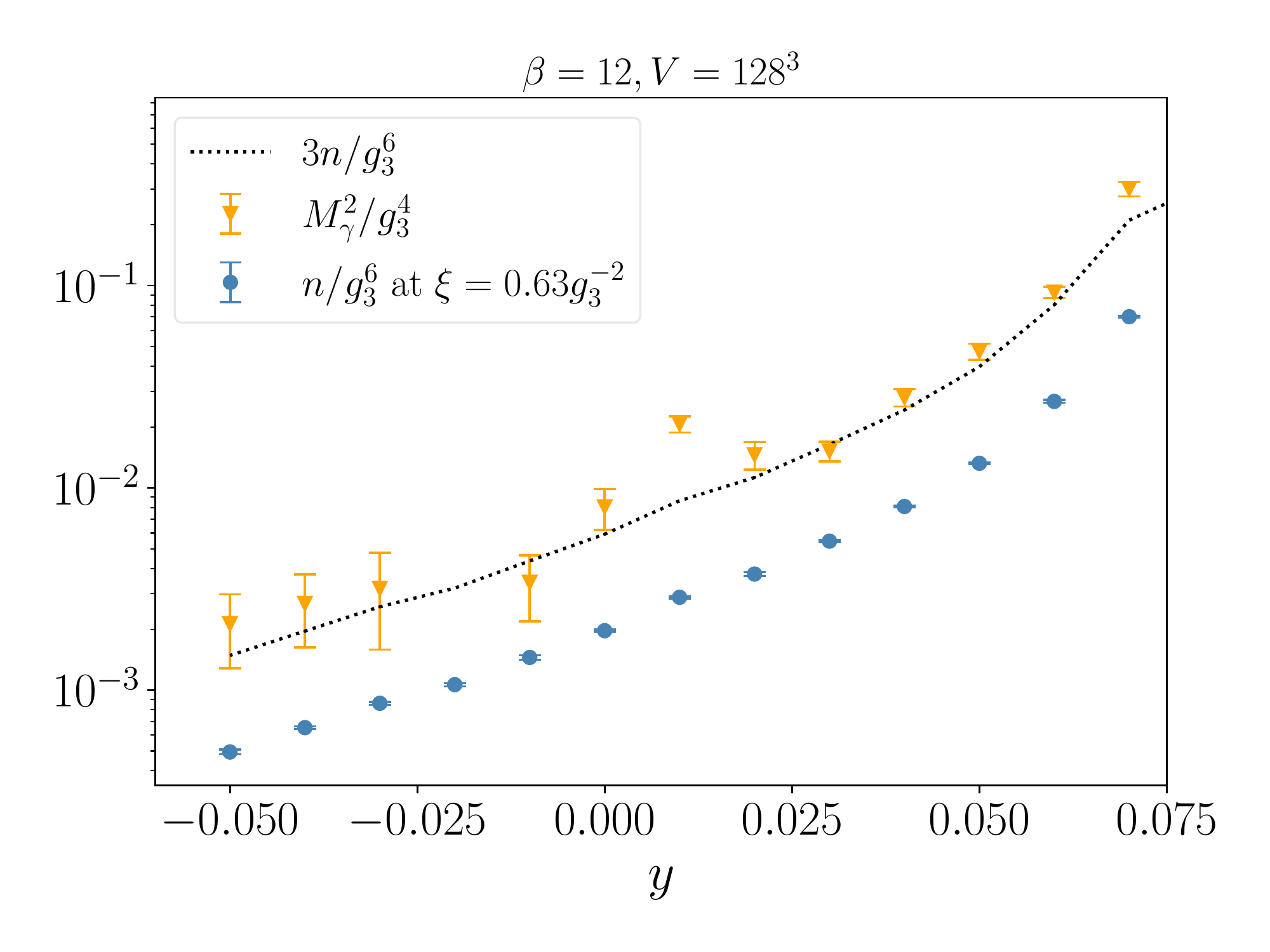}
  \caption{At left, the photon mass squared as a function of $y$; the
    photon becomes almost massless deep in the Higgs phase as expected. At
    right, the squared photon mass is compared to the renormalised
    monopole density, for fixed volume and lattice spacing. \label{fig:photonmass}}
\end{figure}

Our results for the photon mass and its relationship to the monopole
number density are shown in Fig.~\ref{fig:photonmass}.

\section{Conclusions}

We have measured the monopole number density and photon mass in the
$\mathrm{SU}(2)$ Georgi-Glashow model. Our results indicate that
gradient flow provides a meaningful method of renormalising the
monopole number density, giving a result with a well-defined continuum
limit.

In addition, we have used blocked correlators to measure the photon
mass through the crossover. We have established that the semiclassical
proportionality between the monopole number density and the photon
mass~(\ref{eq:semiclassicalrelationship}) holds nonperturbatively, and
furthermore that it extends to the vicinity of the crossover transition 
where the semiclassical arguments are not reliable.

For theories beyond the Standard Model which contain adjoint scalar
fields, the condensation of monopoles naturally plays a role in
crossovers and phase transitions. Our results give qualitative insight
into the infrared behaviour of the Higgs regime near the transition
point. The presence of monopoles could indeed have consequences for
the strength of the transition as previously suggested in
Ref.~\cite{Kajantie:1997tt}. Due to the nonperturbative nature of
processes involving monopoles, caution is warranted when conducting
perturbative studies of such models near the transition point,
particularly for weak first-order transitions.


\begin{thebibliography}{99}


  
\bibitem{Georgi:1974sy}
H.~Georgi and S.~L.~Glashow,
Phys. Rev. Lett. \textbf{32} (1974), 438-441
doi:10.1103/PhysRevLett.32.438

\bibitem{tHooft:1974kcl}
G.~'t Hooft,
Nucl. Phys. B \textbf{79} (1974), 276-284
doi:10.1016/0550-3213(74)90486-6

\bibitem{Polyakov:1974ek}
A.~M.~Polyakov,
JETP Lett. \textbf{20} (1974), 194-195
PRINT-74-1566 (LANDAU-INST).

\bibitem{Polyakov:1976fu}
A.~M.~Polyakov,
Nucl. Phys. B \textbf{120} (1977), 429-458
doi:10.1016/0550-3213(77)90086-4

\bibitem{Patel:2012pi}
H.~H.~Patel and M.~J.~Ramsey-Musolf,
Phys. Rev. D \textbf{88} (2013), 035013
doi:10.1103/PhysRevD.88.035013
[arXiv:1212.5652 [hep-ph]].

\bibitem{Niemi:2020hto}
L.~Niemi, M.~J.~Ramsey-Musolf, T.~V.~I.~Tenkanen and D.~J.~Weir,
Phys. Rev. Lett. \textbf{126} (2021) no.17, 171802
doi:10.1103/PhysRevLett.126.171802
[arXiv:2005.11332 [hep-ph]].

\bibitem{Appelquist:1981vg}
T.~Appelquist and R.~D.~Pisarski,
Phys. Rev. D \textbf{23} (1981), 2305
doi:10.1103/PhysRevD.23.2305

\bibitem{Kajantie:1997tt}
K.~Kajantie, M.~Laine, K.~Rummukainen and M.~E.~Shaposhnikov,
Nucl. Phys. B \textbf{503} (1997), 357-384
doi:10.1016/S0550-3213(97)00425-2
[arXiv:hep-ph/9704416 [hep-ph]].

\bibitem{Davis:2001mg}
A.~C.~Davis, A.~Hart, T.~W.~B.~Kibble and A.~Rajantie,
Phys. Rev. D \textbf{65} (2002), 125008
doi:10.1103/PhysRevD.65.125008
[arXiv:hep-lat/0110154 [hep-lat]].

\bibitem{Luscher:2010iy}
M.~L\"uscher,
JHEP \textbf{08} (2010), 071
[erratum: JHEP \textbf{03} (2014), 092]
doi:10.1007/JHEP08(2010)071
[arXiv:1006.4518 [hep-lat]].

\bibitem{Laine:1995np}
M.~Laine,
Nucl. Phys. B \textbf{451} (1995), 484-504
doi:10.1016/0550-3213(95)00356-W
[arXiv:hep-lat/9504001 [hep-lat]].


\bibitem{Davis:2000kv}
A.~C.~Davis, T.~W.~B.~Kibble, A.~Rajantie and H.~Shanahan,
JHEP \textbf{11} (2000), 010
doi:10.1088/1126-6708/2000/11/010
[arXiv:hep-lat/0009037 [hep-lat]].

\bibitem{Hart:1996ac}
A.~Hart, O.~Philipsen, J.~D.~Stack and M.~Teper,
Phys. Lett. B \textbf{396} (1997), 217-224
doi:10.1016/S0370-2693(97)00104-4
[arXiv:hep-lat/9612021 [hep-lat]].

\end{thebibliography}
\end{document}